\documentclass[prl,twocolumn,showpacs,preprintnumbers,amsmath,amssymb]{revtex4}
\usepackage{graphicx}

\begin{document}

\preprint{APS/123-QED}

\title{Hexaroll chaos in inclined layer convection}

\author{Will Brunner}
 \affiliation{Max Planck Institute for Dynamics and Self Organization \\
 G\"ottingen, Germany}
 
\author{Jonathan McCoy}
\affiliation{Laboratory of Atomic and Solid State Physics, Cornell University, Ithaca, New York
}
\author{Werner Pesch}
 \affiliation{University of Bayreuth, Bayreuth, Germany}

\author{Eberhard Bodenschatz}
 \homepage{http://milou.msc.cornell.edu/}
 \email{Eberhard.Bodenschatz@ds.mpg.de}
\affiliation{Max Planck Institute for Dynamics and Self Organization \\
 G\"ottingen, Germany}
\affiliation{Laboratory of Atomic and Solid State Physics, Cornell University, Ithaca, New York
}
 \date{\today}

\begin{abstract}
We report experimental observations of hexaroll chaos in inclined layer convection. Two separate populations of defect complexes characterize the defect turbulent state, one the classical penta-hepta defect of hexagonal planform patterns, and the other a short-lived complex capable of self-annihilation.  By measuring  the defect statistics we show that short-lived defect complexes give rise to linear defect destruction rates. This observation explains previous results in other defect-turbulent pattern forming systems.  
\end{abstract}

\pacs{47.20.Bp, 47.20.Ky, 47.27.Te, 47.54.+r}
\maketitle
Systems that are driven out of equilibrium often show similar patterns although the underlying processes can be quite different \cite{1993RvMP...65..851C}. One challenge is to find measures  that can    quantitatively assess the similarity of different patterns. A convenient  starting point for a classification  are topological defects,  as has long been known in the field of fingerprint recognition \cite{1892GaltonFingerPrints}.  Much of the previous work  \cite{1990PhRvA..41.1138G,1998PhRvL..81.4120E,2001PhRvL..87q4502G,2002EL.....57..480C,2002EL.....57..824T,2002PhRvL..88c4501D,2003PhRvL..90m4502Y,2003NJPh....5..135Y,2004PhyD..193..208D,2004CHAOS..14..864H,2004CHAOS..14..933W} is concerned with the particle-like appearance of defects and with their chaotic spatiotemporal dynamics, {\it i.e.}, defect turbulence. 

A key feature of defect turbulence is the reaction dynamics between different types of defects.  Gil {\it et al.} \cite{1990PhRvA..41.1138G} showed that the topological property carried by defects is conserved, causing them to behave as charged particles in reactions.  This leads to a generalized Poisson distribution for defect populations, and reaction rates that obey the mass action law, where defects are created with rates independent of $N$, the number of defects present, and are annihilated at a rate proportional to $N^2$.  This result was confirmed \cite{2001PhRvL..87q4502G} in a thermal convection experiment, where they also accounted for defects entering and leaving through the boundaries.  Recently, however, two results have been brought forward that appear to contradict the notion that mass action is followed.  The first of these is a Swift-Hohenberg type model of rotating thermal convection \cite{2003PhRvL..90m4502Y}, and the second is a calculation of the complex Ginzburg-Landau equation with noise added \cite{2004PhRvL..93o4101W}.  Both produce defect statistics where defect annihilation rate has a part proportional to the number of defects present in addition to the expected $N^2$ dependence.  To date an  explanation of this phenomenon is lacking.  

\begin{figure}
\includegraphics[width=86mm]{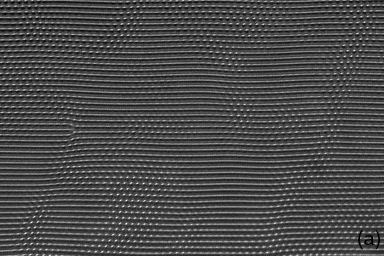}
\includegraphics[width=86mm]{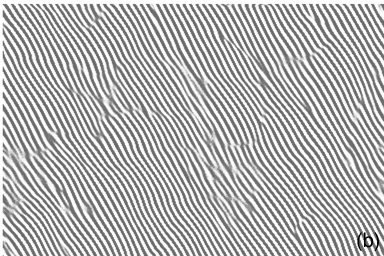}
\caption{\label{sampframe} (a) A sample image from $\epsilon =  0.054$.  Uphill direction is to right.  (b) One oblique mode only,  produced by applying an angular Fourier filter to (a).  Contrast-enhanced for clarity. }
 \end{figure}

Here we  report experimental results from non-Boussinesq inclined layer convection (ILC) where  a inclined layer is heated at one side and  and cooled at the other. It is well known \cite{2000PhRvL..84.5320D} that a shear flow is set up with a flow that is up at the warm side and down at the cold.  For non-Boussinesq conditions in a horizontal layer a hexagonal planform  would be observed \cite{1993RvMP...65..851C}. However, here due to the shear flow one roll orientation is preferred leading to a pattern that is  similar to the hexarolls described for cylindrical rotating convection \cite{1995JFM...301..371A, BrauschThesis}. In our case, we observe a defect turbulent  state of  spatiotemporal chaos (STC) (Fig. \ref{sampframe}) that we term {\it Hexaroll Chaos}. We show that hexaroll chaos  has two separate populations of defect complexes, one the classical penta-hepta defect of hexagonal planform patterns, and the other a short-lived complex capable of self annihilation. By measuring the defect statistics we show that short-lived defect complexes give rise to linear defect destruction rates explaining the earlier observations in other pattern forming systems.
Hexaroll chaos can  be described by amplitude equations of the form
\begin{subequations}
\label{amp}
\begin{align}
\partial_t A_1 = \sigma_1 A_1 - \eta_1 A_2^* A_3^* -A_1 \sum_{j=1}^3 a_{1j}|A_j|^2,\\
\partial_t A_2 = \sigma_2 A_2 - \eta_2 A_1^* A_3^* -A_2 \sum_{j=1}^3 a_{2j}|A_j|^2, \\
\partial_t A_3 = \sigma_3 A_3 - \eta_3 A_1^* A_2^* -A_3 \sum_{j=1}^3 a_{3j}|A_j|^2.
\end{align}
\end{subequations}
Here the convection amplitude is a linear superposition of the three modes: $V({\bf x}) = \sum_{j=1}^3 A_j \exp[i {\bf q}_j {\bf x}] + {\it c.c.}$, where the vectors ${\bf q}$ define a hexagonal lattice, ${\bf q}_1 = (\frac{\sqrt 3}{2}q,-q/2), {\bf q}_2 = (0,q), {\bf q}_3 = (-\frac{\sqrt 3}{2}q,-q/2)$, and the vector ${\bf x}$ represents position.  The convection cell is tilted upward along the ${\bf x}=(1,0)$  direction, breaking the symmetry of the $A_2$ mode, but the other two remain symmetrical, so the following coefficient relations can be inferred:
\begin{equation}
\eta_1 = \eta^*_3, a_{12} =  a^*_{32},  a_{11} =  a^*_{33},  a_{13} =  a^*_{31}
\end{equation}
where a star denotes complex conjugate.  The linear growth rates are given by:
\begin{equation}
\sigma_1 = \epsilon- \Delta \epsilon + i\omega, \sigma_2 = \epsilon , \sigma_3 = \sigma^*_1
\end{equation}
where $\epsilon \equiv \frac{\Delta T-\Delta T_c}{\Delta T_c}$ is the reduced control parameter, $\Delta T$ is the temperature difference driving convection, and $\Delta T_c$ is the value of $\Delta T$ at onset of convection.  $\omega > 0$ represents the downhill drift of modes one and three. The implication of these equations is that convection rolls of mode two appear at $\epsilon = 0$ and the other two modes appear at $\epsilon = \Delta \epsilon$.  We observe $\Delta \epsilon = 1.0 \times 10^{-3} \pm 10^{-4}$ which agrees well with the value of $9.7 \times 10^{-4}$ obtained from numerical solution of the Navier-Stokes equations.  Dislocations primarily arise in modes one and three in regions of low amplitude.  These can be paired into defects of two distinct types.  One type is the classical penta-hepta defect (PHD), where a dislocation of one topological sign in one mode is bound to one of opposite sign in the other mode.  Additionally, we observe a new type of defect complex where short-lived pairs of dislocations of opposite sign are found in the same mode.   We term these dislocation pairs same-mode complexes (SMCs).  

We use  a gas convection apparatus \cite{1996RScI...67.2043D} with a working fluid of CO$_2$ at ($4.240\times 10^6 \pm 10^3$) Pa  regulated to $\pm 300$ Pa and the mean temperature of $(31 \pm 0.05)^\circ$C regulated to $\pm 5 \times 10^{-4} {^\circ}$C. We tilt the plane of the fluid layer to $\theta = 5^\circ$ from the horizontal.  The fluid layer is $d = 308\pm2 \mu$m high, and the vertical diffusion time $\tau_v = 0.56\pm0.01$ s.  Prandtl number is $\sigma  \equiv \kappa/\nu \approx 1$, where $\kappa$ is thermal diffusivity and $\nu$ is kinematic viscosity.  We find onset to rolls at $\Delta T_c = 20.68 \pm 0.01^\circ$C and onset to hexarolls drifting downward with $\omega = 0.8 \tau_v$ s/rad above that at $\Delta \epsilon$.   We choose a region of interest $142d \times 95d$, with the long edge parallel to the direction of tilt, limiting spatial inhomogeneity of $\epsilon$ to $< 10 \%$ during the experiment.  The large value of $\Delta T_c$ makes convection strongly non-Boussinesq, with non-Boussinesq parameters (see, e.g., Ref.~\onlinecite{2003NJPh....5..135Y}) of $\gamma_0 = 0.1653$, $\gamma_1 = -0.2885$, $\gamma_2 = 0.2053$, $\gamma_3 = 0.0085$, and $\gamma_4 = -0.2491$, where $\gamma_0$ and $\gamma_1$ are the first- and second-order coefficients for the dependence of density on temperature, and $\gamma_2$--$\gamma_4$ are the first-order coefficients for the temperature dependence of viscosity, heat conductivity, and specific heat, respectively. 

We collected data by quasistatically  increasing $\epsilon$ in steps of $10^{-3}$ to  $\epsilon =  0.054$, where we acquired $10^6$ shadowgraph images  of the region of interest at 5 frames/second, for an observation time of $357100 \tau_v$.  (See Ref.~\onlinecite{1996RScI...67.2043D} for details of shadowgraphic technique.) An example frame appears in Fig.~\ref{sampframe}.  Note the variation in oblique roll amplitude that gives the appearance of mixed rolls and hexagons, and that oblique-mode dislocations are restricted to those areas of low amplitude.

\begin{figure}
 \includegraphics[width=86mm]{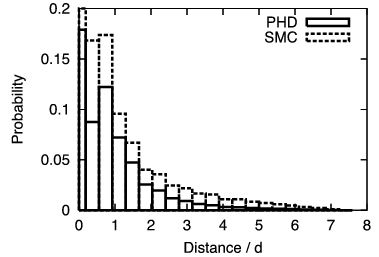}
 \caption{\label{sep} Histogram of separation distance of dislocations in defect complexes}
 \end{figure}

\begin{figure}
\includegraphics[width=86mm]{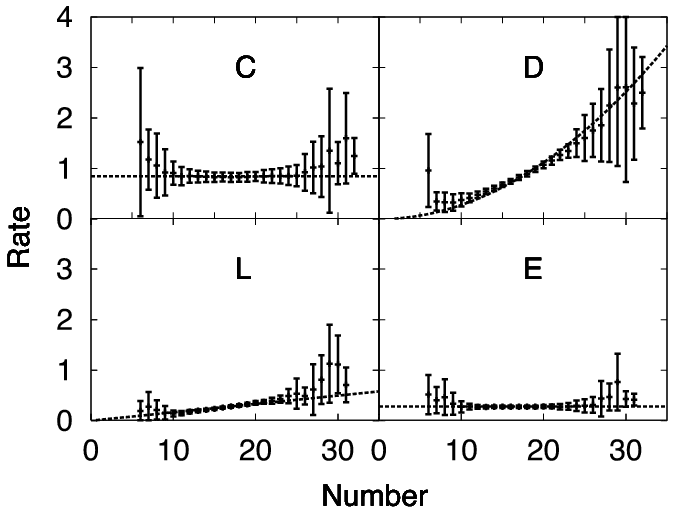}
\caption{\label{phd}PHD Creation, destruction, leaving and entering rates, (C,D,L,E, respectively) along with fits.  Number (N) is the geometric mean of the populations of the two types of PHDs, which are approximately equal.  Rate means number of events for each N, divided by the number of frames containing exactly N.  Error bars represent standard deviation, calculated by subdividing the data set, and are used for weighting fits.  Please note that the mean rate is overestimated when the standard deviation is on the order of the mean because all values are positive.}
\end{figure}

\begin{figure}
\includegraphics[width=86mm]{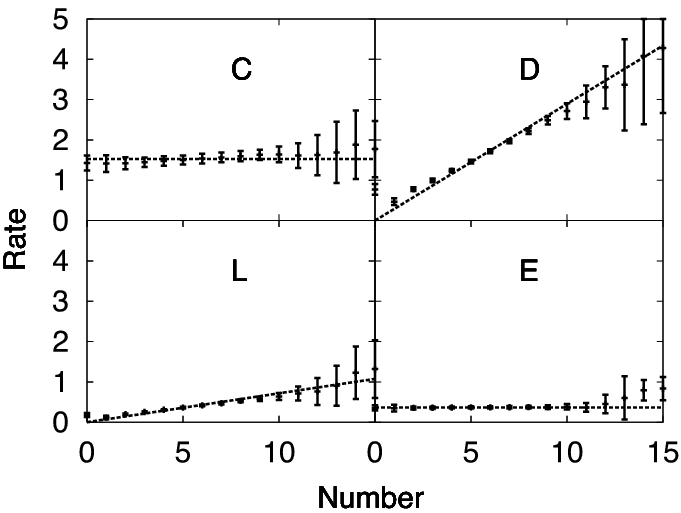}
\caption{\label{smc}SMC creation, destruction, leaving and entering rates (C,D,L,E, respectively).  Number (N) is the geometric mean of the populations of the two types of SMCs}
\end{figure}

 We processed all images by first separating them into three individual modes by applying an angular Fourier filter.  We identified dislocations in each mode using previously described techniques \cite{2002PhRvL..88c4501D}.  Dislocations in mode two (with stripes parallel to tilt) are rare, and were recorded but not processed further.  After all dislocations were identified for all frames, we assigned each to either a PHD or to an SMC.  We did this by iterating through all dislocations, and searching within a threshold radius for an opposite charge dislocation in either the same mode or the other mode.   If such a match was found, both dislocations were marked as being bound, and the search moved on to the unbound dislocations.  After all dislocations were visited, we increased the threshold and repeated the search over all unbound dislocations.  This process was continued until a maximum threshold of $\sim 12d$ is reached (Fig. ~\ref{sep}). Typically, $\sim 90\%$ of dislocations are accounted for in bound pairs, leaving the other $10\%$ unbound at the edges.  PHDs are roughly four times more prevalent than SMCs in all $10^6$ images.  
 The next step of processing was to track the center points of PHDs and SMCs by the dislocation-tracking method of Huepe {\it et al.} \cite{2004CHAOS..14..864H}.  This allowed us to tally defect creation and destruction rates and rates of entering and leaving the frame (Figs.~\ref{phd} \&~\ref{smc}).  We calculated the rate for each N by dividing the number of events by the total number of frames containing N defects, where N is the geometric mean of the two types of either PHDs or SMCs.  We split the data set into 100 equal parts to calculate standard deviation of rates.   

Let us consider the types of reactions that SMCs and PHDs can undergo.  This will illuminate later descriptions of defect creation and destruction rates.  The classical case of a symmetrical hexagonal pattern (see,{\it e.g.,} Ref.~\onlinecite{1995PhRvL..74.4201T}) has six types of PHDs: $1^+2^-$, $1^-2^+$,$1^+3^-$, $1^-3^+$, $2^+3^-$ and $2^-3^+$ where, {\it e.g.}, $1^+$ is a positive dislocation in mode 1.  These follow reactions of the forms
\begin{subequations}
\begin{align}
1^+2^-+1^-3^+&\leftrightarrow 2^-3^+,\\
\label{annphd}
1^+2^-+1^-2^+&\leftrightarrow 0,
\end{align}
\end{subequations}
which yield nine possible reactions once permutations are considered.
Note that all destruction reactions (those which proceed to the right) involve two reactants, so from mass action we expect a rate that depends on the number of PHDs squared.  Creation reactions follow a linear rate law, with a nonzero intercept due to (\ref{annphd}).  Now, if we consider the asymmetric case of hexaroll chaos, we must disallow dislocations in mode two.  There are now two types of PHDs possible: $1^+3^-$ and $1^-3^+$.  In addition, there are two types of SMCs possible: $1^+1^-$ and $3^+3^-$.  The reactions are:
\begin{subequations}
\begin{align}
\label{association}
1^+1^- + 3^+3^- &\leftrightarrow 1^+3^- + 1^-3^+,\\
\label{ann1}
1^+1^- &\leftrightarrow 0,\\
\label{ann2}
3^+3^- &\leftrightarrow 0.
\end{align}
\end{subequations}
Where (\ref{association}) is the realignment of two SMCs into two PHDs, and (\ref{ann1}) and (\ref{ann2}) are the self-annihilation of SMCs.  Here we have two sets of rates, those for PHDs and those for SMCs.  PHDs are created by (\ref{association}) proceeding to the right.  The rate for this reaction is independent  of the number of PHDs because there are no PHDs as reactants, and populations of SMCs and PHDs are uncorrelated.  PHDs are destroyed by the same reaction moving in the opposite direction, with a rate proportional to the number of PHDs squared.  SMCs are created by all three reactions going to the left, which gives a constant rate law.  Finally, SMCs are destroyed by these reactions going to the right, giving a linear rate from (\ref{ann1}) and (\ref{ann2}) with a squared term from (\ref{association}).  In practice, we find that this squared term is negligible, so it is set to zero when constructing fits.  Otherwise,  these rate relations and those for entering and leaving given by Daniels and Bodenschatz \cite{2002PhRvL..88c4501D} fit the data in Figs.~\ref{phd} \&~\ref{smc} with good results.
  

\begin{figure}
\includegraphics[width=86mm]{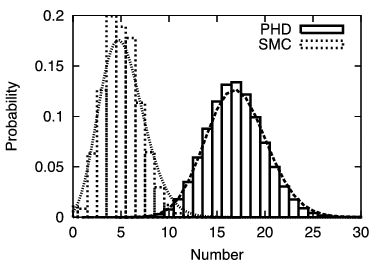}
\caption{\label{prob}Probability distribution functions (PDFs) for average numbers of each type of SMCs and PHDs.  Boxes are measured, and curves are calculated from fits in Figs. \ref{phd} \& \ref{smc} using detailed balance.}
\end{figure}

Using detailed balance, we recursively define a modified Poisson distribution of the probability of finding $N$ defects in the frame at a given time in terms of creation, destruction, leaving, and entering rates:
\begin{equation}
\label{poisson}
\mathcal{P}(N) = \mathcal{P}(0) \prod_{j=1}^N \frac{C(j-1) E(j-1)}{D(j) L(j)}
\end{equation}
where $\mathcal{P}(0)$ is supplied as a normalization condition.  Note that for the case of SMCs where $C$ and $E$ are constants, and $D$ and $L$ are linear functions of $N$, (\ref{poisson}) reduces to the Poission distribution.  For the PHD case, (\ref{poisson}) reduces to the squared Poisson distribution of Gil {\it et al.} \cite{1990PhRvA..41.1138G} with the finite-size extension of Daniels and Bodenschatz \cite{2002PhRvL..88c4501D} (Fig. ~\ref{prob}).
We conclude that defect statistics of hexaroll chaos may be easily understood within the existing framework of reactions and rate laws if we introduce a new type of short-lived defect complexes (SMCs) that usually self-annihilate, but that may stabilize by binding into long-lived PHDs.  These complexes appear to arise because of broken symmetry which changes the allowed ways of creating and destroying PHDs.  We suggest that this may also be the case in other systems, such as the rotating hexagons of Young and Riecke \cite{2003PhRvL..90m4502Y} where their ansatz of mixed linear and quadratic rate laws may be separable into two sets of laws, one for SMCs and one for PHDs.  Similarly, Wang's STC state with noise added  \cite{2004PhRvL..93o4101W} may contain a population of short-lived, noise-induced defects that primarily annihilate with the defect they were created with, thus following a linear rate law.  More detailed study of these systems is called for to determine if there is a generality that connects symmetry breaking with altered rate laws.
\bibliography{paper}
\end{document}